\documentclass[preprint,aps,tightenlines,byrevtex,showpacs,nofootinbib]{revtex4}
\begin{document}

\title{On the exact formula
for neutrino oscillation probability by Kimura, Takamura and Yokomakura}

\author{Osamu Yasuda}
\email{yasuda_at_phys.metro-u.ac.jp}
\affiliation{Department of Physics, Tokyo Metropolitan University,
Minami-Osawa, Hachioji, Tokyo 192-0397, Japan}

\begin{abstract}

The exact formula for the neutrino oscillation probability in matter
with constant density, which was discovered by Kimura, Takamura and
Yokomakura, has been applied mostly to the standard case with
three flavor neutrino so far.  In this paper applications of their formula
to more general cases are discussed.
It is shown that
this formalism can be generalized to various cases where
the matter potential have off-diagonal components, and
the two non-trivial examples are given: the case with
magnetic moments and a magnetic field and the case with non-standard
interactions.
It is pointed out that their formalism can be applied
also to the case in the long baseline limit with matter whose density varies
adiabatically as in the case of solar neutrino.
\end{abstract}
\vskip 0.1cm
\maketitle

\section{introduction}

Neutrino oscillations in
matter~(See, e.g., Ref.~\cite{Bahcall:1995gw} for review.)
have been discussed by many people in the past because
the oscillation probability has non-trivial behaviors in matter
and due to the matter effect it may exhibit
non-trivial enhancement which could be physically important.
Unfortunately,
it is not easy to get an analytical formula
for the oscillation probability in the three flavor
neutrino scheme in matter, and investigation of its behaviors
has been a difficult but important problem in the phenomenology
of neutrino oscillations.
In 2002 Kimura, Takamura and Yokomakura derived a nice compact
formula~\cite{Kimura:2002hb,Kimura:2002wd}
for the neutrino oscillation probability in matter with
constant density.  Basically what they showed is that
the quantity $\tilde{U}^\ast_{\alpha j}\tilde{U}_{\beta j}$,
which is a factor crucial to express the oscillation probability
analytically,
can be expressed as a linear combination of
$U^\ast_{\alpha j}U_{\beta j}$, where $\tilde{U}_{\alpha j}$ and
$U_{\alpha j}$ stand for the matrix element of the
MNS matrix in matter and in vacuum, respectively.

However, their formula is only applicable to the standard
three flavor case.  In this paper we show that their result can be
generalized to various cases.  We also show that their
formalism can be applied also to the case with slowly varying
matter density in the limit of the long neutrino path.
In Sect. \ref{generalities}, we review
briefly some aspects of the oscillation probabilities,
including a simple derivation for the formula by
Kimura, Takamura and Yokomakura which was given in Ref.~\cite{Xing:2005gk},
because these are used in the following sections.
Their formalism is generalized to the various cases where
the matter potential has off-diagonal components,
and we will discuss
the case with large magnetic moments and a magnetic
field (Sect. \ref{magnetic}) and the case with non-standard
interactions (Sect. \ref{nonstandard}).
In Sect. \ref{conclusions} we summarize our conclusions.

\section{generalities about oscillation probabilities \label{generalities}}

\subsection{The case of constant density}
It has been known~\cite{Grimus:1993fz} (See also
earlier works~\cite{Halprin:1986pn,Mannheim:1987ef,Sawyer:1990tw}.)
that after eliminating the negative energy states
by a Tani-Foldy-Wouthusen-type transformation, the
Dirac equation for neutrinos propagating in matter is reduced to the
familiar form:
\begin{eqnarray}
i{d\Psi \over dt}=
\left[U{\cal E}U^{-1}
+{\cal A}(t)
\right]\Psi,
\label{sch1}
\end{eqnarray}
where
\begin{eqnarray}
{\cal E}&\equiv&{\mbox{\rm diag}}\left(E_1,E_2,E_3\right),
\nonumber\\
{\cal A}(t)&\equiv&\sqrt{2}G_F{\mbox{\rm diag}}
\left(N_e(t)-N_n(t)/2,-N_n(t)/2,-N_n(t)/2\right),
\nonumber
\end{eqnarray}
$\Psi^T\equiv(\nu_e,\nu_\mu,\nu_\tau)$ is the flavor eigenstate, $U$
is the Maki-Nakagawa-Sakata (MNS) matrix, $E_j\equiv\sqrt{m_j^2 +{\vec
{\relax{\kern .1em p}}}^2}~(j=1,2,3)$ is the energy eigenvalue of each
mass eigenstate, and the matter effect ${\cal A}(t)$ at time (or
position ) $t$ is characterized by the density $N_e(t)$ of electrons
and the one $N_n(t)$ of neutrons, respectively.
Throughout this paper we assume for simplicity that the density of
matter is either constant or slowly varying so that its derivative is
negligible.
The $3\times3$ matrix on the
right hand side of Eq. (\ref{sch1}) can be formally
diagonalized as:
\begin{eqnarray}
U{\cal E}U^{-1}+{\cal A}(t)
=\tilde{U}(t)\tilde{{\cal E}}(t)\tilde{U}^{-1}(t),
\label{sch3}
\end{eqnarray}
where
\begin{eqnarray}
\tilde{{\cal E}}(t)&\equiv&{\mbox{\rm diag}}\left(
\tilde{E}_1(t),\tilde{E}_2(t),\tilde{E}_3(t)\right)
\nonumber
\end{eqnarray}
is a diagonal matrix with the energy eigenvalues
$\tilde{E}_j(t)$ in the presence of the matter effect.

First of all, let us assume that the matter density
${\cal A}(t)$ is constant.
Then all the $t$ dependence disappears and
Eq. (\ref{sch1}) can be easily solved, resulting
the flavor eigenstate at the distance $L$:
\begin{eqnarray}
\Psi(L)=\tilde{U}\exp\left(-i\tilde{{\cal E}}L\right)\tilde{U}^{-1}\Psi(0).
\label{sol1}
\end{eqnarray}
Thus the oscillation probability
$P(\nu_\alpha\rightarrow\nu_\beta)$ is
given by
\begin{eqnarray}
P(\nu_\alpha\rightarrow\nu_\beta)&=&
\left|\left[\tilde{U}\exp\left(-i{\cal E}L\right)\tilde{U}^{-1}
\right]_{\beta\alpha}\right|^2\nonumber\\
&=&\delta_{\alpha\beta}-4\sum_{j<k}\mbox{\rm Re}\left(\tilde{X}^{\alpha\beta}_j
\tilde{X}^{\alpha\beta\ast}_k\right)
\sin^2\left({\Delta \tilde{E}_{jk}L \over 2}\right)\nonumber\\
&{\ }&+2\sum_{j<k}\mbox{\rm Im}\left(\tilde{X}^{\alpha\beta}_j
\tilde{X}^{\alpha\beta\ast}_k\right)
\sin\left(\Delta \tilde{E}_{jk}L\right),
\label{probv}
\end{eqnarray}
where we have defined
\begin{eqnarray}
\tilde{X}^{\alpha\beta}_j&\equiv&\tilde{U}_{\alpha j}\tilde{U}^\ast_{\beta j},
\nonumber\\
\Delta \tilde{E}_{jk}&\equiv&\tilde{E}_j-\tilde{E}_k,
\nonumber
\end{eqnarray}
and throughout this paper the indices $\alpha, \beta = (e, \mu, \tau)$
and $j, k = (1, 2, 3)$ stand for those of the flavor and mass eigenstates,
respectively.
Once we know the eigenvalues $\tilde{E}_j$ and the
quantity $\tilde{X}^{\alpha\beta}_j$, the oscillation
probability can be expressed analytically.\footnote{
In the standard case with three flavors of neutrinos
in matter, the energy eigenvalues $\tilde{E}_j$ can
be analytically obtained by the root formula
for a cubic equation~\cite{Barger:1980tf}.
So the only non-trivial problem in the standard case
is to obtain the expression for $\tilde{X}^{\alpha\beta}_j$,
and this was done by Kimura, Takamura and
Yokomakura~\cite{Kimura:2002hb,Kimura:2002wd}.
In general cases, however, the analytic expression for
$\tilde{E}_j$ is very difficult or impossible to obtain,
and we will discuss below only examples in which
the analytic expression for $\tilde{E}_j$ is known.}

\subsection{The case of adiabatically varying density}
Secondly, let us consider the case where the density of
the matter varies adiabatically as in the case of
the solar neutrino deficit phenomena.
In this case, instead of Eq. (\ref{sol1}), we get
\begin{eqnarray}
\Psi(L)=\tilde{U}(L)\exp\left[-i\int_0^L\tilde{{\cal E}}(t)\,dt\right]
\tilde{U}(0)^{-1}\Psi(0),
\nonumber
\end{eqnarray}
where $\tilde{U}(0)$ and $\tilde{U}(L)$ stand for
the effective mixing matrices at the origin $t=0$ and
at the end point $t=L$.
The oscillation probability is given by
\begin{eqnarray}
P(\nu_\alpha\rightarrow\nu_\beta)&=&
\left|\left[\tilde{U}(L)\exp\left\{-i\int_0^L\tilde{{\cal E}}(t)\,dt
\right\}\tilde{U}(0)^{-1}
\right]_{\beta\alpha}\right|^2\nonumber\\
&=&
\sum_{j,k}\,\tilde{U}(L)_{\beta j}\tilde{U}(L)^\ast_{\beta k}
\tilde{U}(0)^\ast_{\alpha j}\tilde{U}(0)_{\alpha k}
\exp\left[-i\int_0^L\Delta\tilde{E}(t)_{jk}\,dt\right].
\label{proba1}
\end{eqnarray}
Eq. (\ref{proba1}) requires in general
the quantity like $\tilde{U}(t)_{\beta j}\tilde{U}^\ast(t)_{\beta k}$
which has the same flavor index $\beta$
but different mass eigenstate indices $j, k$,
and it turns out that the analytical expression for
$\tilde{U}(t)_{\beta j}\tilde{U}^\ast(t)_{\beta k}$
is hard to obtain.  However,
if the length $L$ of the neutrino path is very large
and if $|\int_0^L\Delta\tilde{E}(t)_{jk}\,dt|\gg 1$ is satisfied
for $j\ne k$, as in the case of the solar neutrino deficit phenomena,
after averaging over rapid oscillations
Eq. (\ref{proba1}) is reduced to
\begin{eqnarray}
P(\nu_\alpha\rightarrow\nu_\beta)&=&
\sum_j\,\tilde{X}^{\beta\beta}_j(L)\tilde{X}^{\alpha\alpha}_j(0),
\nonumber
\end{eqnarray}
where we have defined
\begin{eqnarray}
\tilde{X}^{\alpha\alpha}_j(t)\equiv
\left|\tilde{U}(t)_{\alpha j}\right|^2.
\nonumber
\end{eqnarray}
In the case of the solar neutrinos deficit process
$\nu_e\rightarrow\nu_e$
during the daylight,
$\tilde{X}^{\beta\beta}_j(L)$ at the end point $t=L$
and $\tilde{X}^{\alpha\alpha}_j(0)$ at the origin $t=0$
correspond to $X^{\beta\beta}_j$ in vacuum and
$[\tilde{X}^{\alpha\alpha}_j]_\odot$
at the center of the Sun, respectively, where
\begin{eqnarray}
X^{\alpha\beta}_j&\equiv& U_{\alpha j}U^\ast_{\beta j}
\nonumber\\
\left[\tilde{X}^{\alpha\beta}_j\right]_\odot
&\equiv&
\left[\tilde{U}_{\alpha j}\tilde{U}^\ast_{\beta j}\right]_\odot
\nonumber
\end{eqnarray}
are bilinear products of the elements of the mixing matrices
in vacuum and at the center of the Sun,
respectively.  Thus we obtain
\begin{eqnarray}
P(\nu_e\rightarrow\nu_e)
&=&\sum_j\,X^{ee}_j
\left[\tilde{X}^{ee}_j\right]_\odot.
\nonumber
\end{eqnarray}
Hence we see that evaluation of the quantity
$\tilde{X}^{\alpha\alpha}_j$ in the presence of the matter
effect is important
not only in the case of constant matter density
but also in the case of adiabatically varying density.

\subsection{Another derivation of the formula by
Kimura, Takamura and Yokomakura \label{standard}}
In this subsection a systematic derivation of their formula
is given because such a derivation will be crucial
for the generalizations in the following sections.\footnote{
The argument here is
the same as that in Ref.~\cite{Xing:2005gk}.
Since this derivation does not seem to be widely
known, it is reviewed here.}
The arguments are based on the trivial identities.
From the unitarity condition of the matrix $\tilde{U}$,
we have
\begin{eqnarray}
\delta_{\alpha\beta}=\left[\tilde{U}\tilde{U}^{-1}\right]_{\alpha\beta}
=\sum_j\tilde{U}_{\alpha j}\tilde{U}^\ast_{\beta j}
=\sum_j\tilde{X}^{\alpha\beta}_j.
\label{const1}
\end{eqnarray}
Next we take the $(\alpha,\beta)$ component of the both hand sides
in Eq. (\ref{sch3}):
\begin{eqnarray}
\left[U{\cal E}U^{-1}+{\cal A}\right]_{\alpha\beta}
=\left[\tilde{U}\tilde{{\cal E}}\tilde{U}^{-1}\right]_{\alpha\beta}
=\sum_j\tilde{U}_{\alpha j}\tilde{E}_j\tilde{U}^\ast_{\beta j}
=\sum_j\tilde{E}_j\tilde{X}^{\alpha\beta}_j
\label{const2}
\end{eqnarray}
Furthermore, we take the $(\alpha,\beta)$ component of
the square of Eq. (\ref{sch3}):
\begin{eqnarray}
\left[\left(U{\cal E}U^{-1}+{\cal A}\right)^2\right]_{\alpha\beta}
=\left[\tilde{U}\tilde{{\cal E}}^2\tilde{U}^{-1}\right]_{\alpha\beta}
=\sum_j\tilde{U}_{\alpha j}\tilde{E}^2_j\tilde{U}^\ast_{\beta j}
=\sum_j\tilde{E}^2_j\tilde{X}^{\alpha\beta}_j
\label{const3}
\end{eqnarray}
Putting Eqs. (\ref{const1})--(\ref{const3}) together,
we have
\begin{eqnarray}
\left(\begin{array}{ccc}
1&1&1\cr
\tilde{E}_1&\tilde{E}_2&\tilde{E}_3\cr
\tilde{E}^2_1&\tilde{E}^2_2&\tilde{E}^2_3
\end{array}\right)
\left(\begin{array}{c}
\tilde{X}^{\alpha\beta}_1\cr
\tilde{X}^{\alpha\beta}_2\cr
\tilde{X}^{\alpha\beta}_3
\end{array}\right)
=\left(\begin{array}{r}
\delta_{\alpha\beta}\cr
\left[U{\cal E}U^{-1}+{\cal A}\right]_{\alpha\beta}\cr
\left[\left(U{\cal E}U^{-1}+{\cal A}\right)^2\right]_{\alpha\beta}
\end{array}\right),
\nonumber
\end{eqnarray}
which can be easily solved by inverting the
Vandermonde matrix:
\begin{eqnarray}
\left(\begin{array}{c}
\tilde{X}^{\alpha\beta}_1\cr
\tilde{X}^{\alpha\beta}_2\cr
\tilde{X}^{\alpha\beta}_3
\end{array}\right)
=\left(\begin{array}{ccc}
\displaystyle
\frac{{\ }1}{\Delta \tilde{E}_{21} \Delta \tilde{E}_{31}}
(\tilde{E}_2\tilde{E}_3, & -(\tilde{E}_2+\tilde{E}_3),&
1)\cr
\displaystyle
\frac{-1}{\Delta \tilde{E}_{21} \Delta \tilde{E}_{32}}
(\tilde{E}_3\tilde{E}_1, & -(\tilde{E}_3+\tilde{E}_1),&
1)\cr
\displaystyle
\frac{{\ }1}{\Delta \tilde{E}_{31} \Delta \tilde{E}_{32}}
(\tilde{E}_1\tilde{E}_2, & -(\tilde{E}_1+\tilde{E}_2),&
1)\cr
\end{array}\right)
\left(\begin{array}{r}
\delta_{\alpha\beta}\cr
\left[U{\cal E}U^{-1}+{\cal A}\right]_{\alpha\beta}\cr
\left[\left(U{\cal E}U^{-1}+{\cal A}\right)^2\right]_{\alpha\beta}
\end{array}\right).
\label{solx}
\end{eqnarray}
$[(U{\cal E}U^{-1}+{\cal A})^j]_{\alpha\beta}$ $(j=1,2)$ on
the right hand side are given by the known quantities:
\begin{eqnarray}
\left[U{\cal E}U^{-1}+{\cal A}\right]_{\alpha\beta}&=&
\sum_j E_j X^{\alpha\beta}_j+ A\,\delta_{{\alpha}e}\delta_{{\beta}e}\nonumber\\
\left[\left(U{\cal E}U^{-1}+{\cal A}\right)^2\right]_{\alpha\beta}&=&
\sum_j E^2_j X^{\alpha\beta}_j
+A \sum_j E_j \left(\delta_{{\alpha}e}X^{e\beta}_j
+\delta_{{\beta}e}X^{{\alpha}e}_j\right)
+ A^2\,\delta_{{\alpha}e}\delta_{{\beta}e}.
\nonumber
\end{eqnarray}
It can be shown that Eq. (\ref{solx}) coincides with the
original results by Kimura, Takamura and Yokomakura
\cite{Kimura:2002hb,Kimura:2002wd}.

A remark is in order on Eq. (\ref{solx}).
Addition of a matrix $c{\bf 1}$ to
Eq. (\ref{sch3}) where $c$ is a constant
and ${\bf 1}$ is the identity matrix,
or in other words, the shift
\begin{eqnarray}
E_j \rightarrow E_j + c~~(j=1,2,3),
\label{shift}
\end{eqnarray}
should give the same result for
$\tilde{X}^{\alpha\beta}_j~(j=1,2,3)$,
since Eq. (\ref{shift}) only affects the overall phase of
the oscillation amplitude and the phase has to disappear
in the probability.
It is easy to show that the shift (\ref{shift}) indeed
gives the same result as Eq. (\ref{solx}).  The proof is given in
Appendix \ref{appendix1}.
In practical calculations below, we will always put $c=-E_1$, i.e.,
we will consider the mass matrix $U({\cal E}-E_1\mbox{\bf 1})U^{-1}+{\cal A}$
instead of the original one $U{\cal E}U^{-1}+{\cal A}$, since
all the diagonal elements
$({\cal E}-E_1\mbox{\bf 1})_{jj}=\Delta E_{j1}=\Delta m^2_{j1}/2E$ are expressed
in terms of the relevant variables $\Delta m^2_{j1}$, and therefore
calculations become simpler.  To save space, however, we will
use the matrix $U{\cal E}U^{-1}+{\cal A}$ in most of the following discussions.

\subsection{The case with arbitrary number of neutrinos \label{sterile}}

It is straightforward to generalize the discussions
in sect. \ref{standard} to the case with arbitrary number of neutrinos
where the matter potential is diagonal in the flavor eigenstate.
The scheme with number of sterile neutrinos is one of the 
example of these cases~\cite{Xing:2005gk,Zhang:2006yq}.
The time evolution of such a scheme with $N$ neutrino flavors
is described by
\begin{eqnarray}
i{d\Psi_N \over dt}=
\left(U_N{\cal E}_NU^{-1}_N+{\cal A}_N\right)\Psi_N,
\nonumber
\end{eqnarray}
where
$\Psi^T_N\equiv(\nu_{\alpha_1},\nu_{\alpha_2},\cdots,\nu_{\alpha_N})$
is the flavor eigenstate,
\begin{eqnarray}
{\cal E}_N&\equiv&{\mbox{\rm diag}}\left(E_1,E_2,\cdots,E_N\right)\\
\nonumber
\end{eqnarray}
is the energy matrix of the mass eigenstate,
\begin{eqnarray}
{\cal A}_N&\equiv&{\mbox{\rm diag}}\left(A_1,A_2,\cdots,A_N\right),
\nonumber
\end{eqnarray}
is the potential matrix for the flavor eigenstate,
and $U_N$ is the $N\times N$ MNS matrix.
As in the previous sect., by taking the $\alpha, \beta$
components, we get
\begin{eqnarray}
\sum_j\tilde{E}^m_j\tilde{X}^{\alpha\beta}_j
=\left[\left(U_N{\cal E}_NU^{-1}_N+{\cal A}_N\right)^m\right]_{\alpha\beta}
\quad\mbox{\rm for}~m=0,\cdots,N-1,
\nonumber
\end{eqnarray}
which leads to the simultaneous equation
\begin{eqnarray}
\left(\begin{array}{llll}
1&1&\cdots&1\cr
\tilde{E}_1&\tilde{E}_2&\cdots&\tilde{E}_N\cr
\vdots&\vdots&&\vdots\cr
\tilde{E}^{N-1}_1&\tilde{E}^{N-1}_2&\cdots&\tilde{E}^{N-1}_N
\end{array}\right)
\left(\begin{array}{c}
\tilde{X}^{\alpha\beta}_1\cr
\tilde{X}^{\alpha\beta}_2\cr
\vdots\cr
\tilde{X}^{\alpha\beta}_N
\end{array}\right)
=\left(\begin{array}{c}
\delta_{\alpha\beta}\cr
\left[U_N{\cal E}U^{-1}_N+{\cal A}_N\right]_{\alpha\beta}\cr
\vdots\cr
\left[\left(U_N{\cal E}U^{-1}_N+{\cal A}_N\right)^{N-1}\right]_{\alpha\beta}
\end{array}\right).
\label{simultaneous1}
\end{eqnarray}
Eq. (\ref{simultaneous1}) can be solved by inverting the
$N\times N$ Vandermonde matrix $V_N$:
\begin{eqnarray}
\left(\begin{array}{c}
\tilde{X}^{\alpha\beta}_1\cr
\tilde{X}^{\alpha\beta}_2\cr
\vdots\cr
\tilde{X}^{\alpha\beta}_N
\end{array}\right)
=~V^{-1}_N~
\left(\begin{array}{c}
\delta_{\alpha\beta}\cr
\left[U_N{\cal E}_NU^{-1}_N+{\cal A}_N\right]_{\alpha\beta}\cr
\vdots\cr
\left[\left(U_N{\cal E}_NU^{-1}_N+{\cal A}_N\right)^{N-1}\right]_{\alpha\beta}
\end{array}\right).
\label{solxn}
\end{eqnarray}
The determinant of $V_N$ is the Vandermonde determinant
$\prod_{j<k}\,\Delta\tilde{E}_{jk}$, and therefore $V^{-1}$ can be
analytically obtained as long as we know the value of $\tilde{E}_j$.
The factors $[(U_N{\cal E}_NU^{-1}_N+{\cal A}_N)^j]_{\alpha\beta}$
on the right hand side of Eq. (\ref{solxn})
can be expressed as functions of the energy $E_j$,
the quantity $X^{\alpha\beta}_j$
in vacuum and the matter potential $A_\gamma$, since
the matrix $(U_N{\cal E}_NU^{-1}_N+{\cal A}_N)^j$ is a sum of
products of the matrices $[(U_N{\cal E}_NU^{-1}_N)^\ell]_{\gamma\delta}=
\sum_k E_j^\ell X^{\gamma\delta}_k$ $(0\le \ell\le j)$ and
$[({\cal A}_N)^m]_{\epsilon\eta}=(A_\epsilon)^m\delta_{\epsilon\eta}$
$(0\le m\le j)$.
From Eq. (\ref{solxn}) it is clear that enhancement of the oscillation
probability due to the matter effect occurs only when
some of $\Delta\tilde{E}_{jk}$ becomes small.

\section{the case with large magnetic moments and a magnetic
field \label{magnetic}}
So far we have assumed that the potential term is diagonal
in the flavor basis.
We can generalize the present result to the cases
where we have off-diagonal potential terms.
One of such examples is the case where
there are only three active neutrinos with
magnetic moments and the magnetic field~(See, e.g.,
Ref.~\cite{Bahcall:1995gw} for review.).
The hermitian matrix\footnote{
See~\cite{Grimus:1993fz} for derivation of Eq. (\ref{matrixb})
from the Dirac Eq.}
\begin{eqnarray}
{\cal M}\equiv
\left(\begin{array}{cc}
U{\cal E}U^{-1}&{\cal B}\cr
{\cal B}^\dagger&U^\ast{\cal E}(U^\ast)^{-1}
\end{array}\right)
\label{matrixb}
\end{eqnarray}
with
\begin{eqnarray}
{\cal B}\equiv B\,\mu_{\alpha\beta}
\nonumber
\end{eqnarray}
is the mass matrix
for neutrinos and anti-neutrinos without the matter effect
where neutrinos have the magnetic moments $\mu_{\alpha\beta}$
in the magnetic field $B$.
Here we assume the magnetic interaction of Majorana type
\begin{eqnarray}
\mu_{\alpha\beta}\bar{\nu}_\alpha\,F_{\lambda\kappa}
\sigma^{\lambda\kappa}\,\nu^c_\beta + h. c.,
\label{majorana}
\end{eqnarray}
and in this case the magnetic moments $\mu_{\alpha\beta}$
are real and anti-symmetric in flavor
indices: $\mu_{\alpha\beta}=-\mu_{\beta\alpha}$.

If the magnetic field is constant, then
the oscillation probability can be written as
\begin{eqnarray}
P(\nu_A\rightarrow\nu_B)&=&
\delta_{AB}-4\sum_{J<K}\mbox{\rm Re}\left(\tilde{X}^{AB}_J
\tilde{X}^{AB\ast}_K\right)
\sin^2\left({\Delta \tilde{E}_{JK}L \over 2}\right)\nonumber\\
&{\ }&+2\sum_{J<K}\mbox{\rm Im}\left(\tilde{X}^{AB}_J
\tilde{X}^{AB\ast}_K\right)
\sin\left(\Delta \tilde{E}_{JK}L\right),
\label{probvbc}
\end{eqnarray}
where $A, B$ run $e, \mu, \tau, \bar{e}, \bar{\mu}, \bar{\tau}$,
and $J, K$ run 1, $\cdots$, 6, respectively, and
$\tilde{X}^{AB}_J\equiv U_{AJ}U^\ast_{BJ}$.
$\tilde{E}_J$ ($J=1,\cdots,6$) are the eigenvalues of
the $6\times6$ matrix ${\cal M}$.
On the other hand, if the magnetic field
varies very slowly and if the length $L$ of the baseline
is so long that $|\Delta \tilde{E}_{JK}L|\gg 1$ is satisfied
for $J\ne K$, then the oscillation probability is
given by
\begin{eqnarray}
P(\nu_A\rightarrow\nu_B)&=&
\sum_{J=1}^6\tilde{X}^{BB}_J(L)
\tilde{X}^{AA}_J(0).
\label{probvba}
\end{eqnarray}
Following the same arguments as before,
the quantity $\tilde{X}^{AB}_J$ is given by
inverting the $6\times6$ Vandermonde matrix $V_6$:
\begin{eqnarray}
\left(\begin{array}{c}
\tilde{X}^{AB}_1\cr
\tilde{X}^{AB}_2\cr
\vdots\cr
\tilde{X}^{AB}_6
\end{array}\right)
=~V^{-1}_6~
\left(\begin{array}{c}
\delta_{AB}\cr
\left[{\cal M}\right]_{AB}\cr
\vdots\cr
\left[\left({\cal M}\right)^5\right]_{AB}
\end{array}\right).
\label{solxb}
\end{eqnarray}
As in the previous sections,
$[({\cal M})^J]_{AB}$ $(J=0,\cdots,5)$ on the right hand side of
Eq. (\ref{solxb}) can be expressed in terms of the known quantities
$X^{AB}_K$ and ${\cal B}_{CD}$, and
Eqs. (\ref{probvbc}) and (\ref{solxb}) are useful
only when we know the eigenvalues $\tilde{E}_J$.

To demonstrate the usefulness of these formulae,
let us consider the case where the magnetic field is
large at origin but is zero at the end point and
the magnetic field varies adiabatically.
For simplicity we assume that $\theta_{13}$
and all the CP phases vanish.\footnote{
In the presence of the magnetic interaction (\ref{majorana})
of Majorana type,
the two CP phases, which are absorbed by
redefinition of the charged lepton fields in the
standard case, cannot be absorbed and therefore
become physical.  Here, however, we will assume for simplicity
that these CP phases vanish.}
In this case the $6\times6$ matrix ${\cal M}$ in Eq. (\ref{matrixb})
becomes real, and we obtain the following oscillation probabilities:
\begin{eqnarray}
P(\nu_\alpha\rightarrow\nu_\beta)&=&
P(\bar{\nu}_\alpha\rightarrow\bar{\nu}_\beta)
=\sum_{j=1}^3\,(U_{\beta j})^2[\mbox{\rm Re}\,\tilde{U}(0)_{\alpha j}]^2
\nonumber\\
P(\nu_\alpha\rightarrow\bar{\nu}_\beta)&=&
P(\bar{\nu}_\alpha\rightarrow\nu_\beta)
=\sum_{j=1}^3\,(U_{\beta j})^2[\mbox{\rm Im}\,\tilde{U}(0)_{\alpha j}]^2,
\label{probb}
\end{eqnarray}
where
$\tilde{U}(0)$ the $3\times3$ unitary matrix which diagonalizes
the $3\times3$ matrix $U{\cal E}U^{-1}+i{\cal B}(0)$ at the origin:
\begin{eqnarray}
U{\cal E}U^{-1}+i{\cal B}(0)
=\tilde{U}(0)\tilde{{\cal E}}(0)\tilde{U}^{-1}(0).
\nonumber
\end{eqnarray}
In this example the energy eigenvalues are degenerate, i.e.,
the $6\times6$ energy matrix becomes diag$(\tilde{{\cal E}},\tilde{{\cal E}})$,
and the oscillation probability differs from Eq. (\ref{probvba})
because the condition $|\Delta \tilde{E}_{JK}L|\gg 1 (J\ne K)$
is not satisfied (e.g., $\Delta\tilde{E}_{JK}=0$ not only for
$J=K=1$ but also for $J=1$, $K=4$).
Each probability in Eqs. (\ref{probb}) itself is not expressed in
terms of $\tilde{X}^{\alpha\alpha}_j(0)$, but we find that the
following relation holds:
\begin{eqnarray}
P(\nu_\alpha\rightarrow\nu_\beta)+P(\bar{\nu}_\alpha\rightarrow\nu_\beta)
=\sum_{j=1}^3\,(U_{\beta j})^2|\tilde{U}(0)_{\alpha j}|^2
=\sum_{j=1}^3\,X^{\beta\beta}_j\tilde{X}^{\alpha\alpha}_j(0).
\label{relationb}
\end{eqnarray}
Eq. (\ref{relationb}) is a new result and
without the present formalism it would be
hard to derive it.
The details of derivation of Eq. (\ref{probb}) and explicit
forms of $\tilde{X}^{\alpha\alpha}_j(0)$
are given in Appendix \ref{appendix2}.
Eq. (\ref{relationb}) may be applicable to the case
where high energy astrophysical neutrinos,
which are produced in a relatively large magnetic field,
are observed on the Earth, on the assumption that the fluxes of
neutrinos and anti-neutrinos are almost equal.

\section{the case with non-standard interactions \label{nonstandard}}

Another interesting application is the oscillation probability
in the presence of new physics in propagation~\cite{Guzzo:1991hi,Roulet:1991sm}.
In this case the mass matrix is given by
\begin{eqnarray}
U{\cal E}U^{-1}
+{\cal A}_{NP}
\label{matrixnp}
\end{eqnarray}
where
\begin{eqnarray}
{\cal A}_{NP}&\equiv&\sqrt{2}G_FN_e
\left(
\begin{array}{ccc}
 1+\epsilon_{ee} & \epsilon_{e\mu} & \epsilon_{e\tau}\\
 \epsilon_{e\mu}^\ast & \epsilon_{\mu\mu} & \epsilon_{\mu\tau}\\
 \epsilon_{e\tau}^\ast & \epsilon_{\mu\tau}^\ast & \epsilon_{\tau\tau}
\end{array}
\right).
\nonumber
\end{eqnarray}
The dimensionless quantities
$\epsilon_{\alpha\beta}$ stand for possible deviation from
the standard matter effect.
Also in this case the oscillation probability
is given by Eqs. (\ref{probv}) and (\ref{solx}),
where the standard potential matrix ${\cal A}$
has to be replaced by ${\cal A}_{NP}$.
The extra complication compared to the standard case
is calculations of the eigenvalues $\tilde{E}_j$
and the elements $[(U{\cal E}U^{-1}+{\cal A}_{NP})^m]_{\alpha\beta}$
($m=1,2$).

Again to demonstrate the usefulness of the formalism,
here we will discuss for simplicity the case in which the eigenvalues
are the roots of a quadratic equation.
It is known~\cite{Davidson:2003ha}
that the constraints on the three parameters
$\epsilon_{ee}, \epsilon_{e\tau}, \epsilon_{\tau\tau}$
from various experimental data are weak and they
could be as large as ${\cal O}(1)$.
In Ref.~\cite{Friedland:2005vy} it was found that large values
($\sim {\cal O}(1)$) of the
parameters $\epsilon_{ee}, \epsilon_{e\tau}, \epsilon_{\tau\tau}$ are
consistent with all the experimental data including those of the
atmospheric neutrino data, provided that one of the
eigenvalues of the matrix (\ref{matrixnp}) at high energy limit,
i.e., ${\cal A}_{NP}$,
becomes zero.
Simplifying even further, here we will neglect the
parameters $\epsilon_{e\mu}$, $\epsilon_{\mu\mu}$,
$\epsilon_{\mu\tau}$ which are smaller than ${\cal O}(10^{-2})$
and we will consider the potential matrix
\begin{eqnarray}
{\cal A}_{NP}&=&A
\left(
\begin{array}{ccc}
 1+\epsilon_{ee} & 0 & \epsilon_{e\tau}\\
 0 & 0 & 0\\
 \epsilon_{e\tau}^\ast & 0 & \epsilon_{\tau\tau}
\end{array}
\right),
\label{potentialnp}
\end{eqnarray}
where $A\equiv\sqrt{2}G_FN_e$, the three parameters $\epsilon_{ee}$,
$\epsilon_{e\tau}$, $\epsilon_{\tau\tau}$
are constrained in such a way that
two of the three eigenvalues become zero.
We will assume that $N_e$ is constant,
and we will take the limit $\Delta m^2_{21}\rightarrow0$.
The oscillation probability $P(\nu_\mu\rightarrow\nu_e)$
in this case can be analytically expressed and is given by
\begin{eqnarray}
P(\nu_\mu\rightarrow\nu_e)&=&
-4\mbox{\rm Re}\left(\tilde{X}^{\mu e}_1\tilde{X}^{\mu e\ast}_2\right)
\sin^2\left(\frac{\Lambda_-L}{2}\right)
-4\mbox{\rm Re}\left(\tilde{X}^{\mu e}_2\tilde{X}^{\mu e\ast}_3\right)
\sin^2\left(\frac{\Lambda_+L}{2}\right)
\nonumber\\
&{\ }&-4\mbox{\rm Re}\left(\tilde{X}^{\mu e}_1\tilde{X}^{\mu e\ast}_3\right)
\sin^2\left[\frac{(\Lambda_+-\Lambda_-L)L}{2}\right]\nonumber\\
&{\ }&+\frac{8A(\Delta E_{31})^2}{\Lambda_+\Lambda_-(\Lambda_+-\Lambda_-)}
\,\left|\epsilon_{e\tau}X^{e\mu}_3X^{\mu \tau}_3\right|\,
\sin(\mbox{\rm arg}(\epsilon_{e\mu})+\delta)\nonumber\\
&{\ }&\times
\sin\left(\frac{\Lambda_-L}{2}\right)\sin\left(\frac{\Lambda_+L}{2}\right)
\sin\left[\frac{(\Lambda_+-\Lambda_-)L}{2}\right].
\label{probnp}
\end{eqnarray}
Eq. (\ref{probnp}) is another new result and it would be
difficult to obtain it
without using the present formalism.
The details of derivation of Eq. (\ref{probnp}), explanation of the notations
and the explicit forms of all the variables in Eq. (\ref{probnp}) 
are described in Appendix \ref{appendix3}.

\section{conclusions \label{conclusions}}
The essence of the exact formula for
the neutrino oscillation probability in constant matter
which was discovered by Kimura, Takamura and
Yokomakura lies in the fact that the combination
$\tilde{X}^{\alpha\beta}_j\equiv
\tilde{U}^{\alpha j}\tilde{U}^{\beta j\ast}$ of
the mixing matrix elements in matter can be
expressed as polynomials in the same quantity
$X^{\alpha\beta}_j\equiv
U^{\alpha j}U^{\beta j\ast}$ in vacuum.
In this paper we have discussed applications of their formalism to
more general cases.
We have pointed out that their formalism can be useful
for the cases in matter not only with constant density
but also with density which varies
adiabatically as in the case of the solar neutrino problem,
after taking the limit of the long neutrino path.
We have shown that their formalism can be
generalized to the cases
where the matter potential has off-diagonal components.
As concrete non-trivial examples, we discussed the case with
magnetic moments and a magnetic field, and the case with non-standard
interactions.
The application of the present formalism to the case
with unitarity violation has been discussed
elsewhere~\cite{Fernandez-Martinez:2007ms}.
The formalism by Kimura, Takamura and
Yokomakura is quite general and can be applicable
to many problems in neutrino oscillation phenomenology.

\appendix
\section{proof that Eq. (\ref{shift}) gives the same
(\ref{solx})\label{appendix1}}
In this appendix we show that Eq. (\ref{shift}) gives
the same result for $\tilde{X}^{\alpha\beta}_j~(j=1,2,3)$.
The value of
$\tilde{X}^{\alpha\beta}_j~(j=1,2,3)$ for 
\begin{eqnarray}
\tilde{U}\left(\tilde{{\cal E}}+c\mbox{\bf 1}\right)
\tilde{U}^{-1}=U{\cal E}U^{-1}+{\cal A}+c\mbox{\bf 1} 
\nonumber
\end{eqnarray}
becomes at most quadratic\footnote{
Notice that all the factors $\Delta \tilde{E}_{jk}$
are invariant under the shift (\ref{shift}), and
the only change by this shift comes either from the terms
$\tilde{E}_j\tilde{E}_k$ or from $\tilde{E}_j+\tilde{E}_k$
in the inverse of the Vandermonde matrix (cf. Eq. (\ref{solx})).
Hence the difference by Eq. (\ref{shift}) is at most quadratic
in $c$.} in $c$,
and all one has to do is to show
that the coefficients of the terms linear and quadratic
in $c$ vanish.
Let us introduce the notation
\begin{eqnarray}
\left(\begin{array}{ccc}
1&1&1\cr
\tilde{E}_1+c&\tilde{E}_2+c&\tilde{E}_3+c\cr
(\tilde{E}_1+c)^2&(\tilde{E}_2+c)^2&(\tilde{E}_3+c)^2
\end{array}\right)^{-1}
&\equiv& (V^{-1})^{(0)}+c(V^{-1})^{(1)}+c^2(V^{-1})^{(2)}
\nonumber\cr
\left(\begin{array}{r}
\delta_{\alpha\beta}\cr
\left[U{\cal E}U^{-1}+{\cal A}+c\mbox{\bf 1} \right]_{\alpha\beta}\cr
\left[\left(U{\cal E}U^{-1}+{\cal A}+c\mbox{\bf 1}
\right)^2\right]_{\alpha\beta}
\end{array}\right)
&\equiv& \vec{B}^{(0)}+c\vec{B}^{(1)}+c^2\vec{B}^{(2)},
\nonumber
\end{eqnarray}
where $V^{(k)}$ is the coefficient of the
inverted Vandermonde matrix which is $k$-th order in $c$,
and $B_j^{(k)}$ is the coefficient of the
vector $\left(U{\cal E}U^{-1}+{\cal A}+c{\bf 1}\right)^j$
which is $k$-th order in $c$.
Then the terms linear in $c$ are given by
\begin{eqnarray}
&{\ }&(V^{-1})^{(1)}\vec{B}^{(0)}+(V^{-1})^{(0)}\vec{B}^{(1)}\nonumber\cr
&=&
\left(\begin{array}{ccc}
\displaystyle
\frac{1}{\Delta \tilde{E}_{21} \Delta \tilde{E}_{31}}
(\tilde{E}_2+\tilde{E}_3, & -2,&0)\cr
\displaystyle
\frac{1}{\Delta \tilde{E}_{21} \Delta \tilde{E}_{32}}
(-(\tilde{E}_3+\tilde{E}_1), & +2,&0)\cr
\displaystyle
\frac{1}{\Delta \tilde{E}_{31} \Delta \tilde{E}_{32}}
(\tilde{E}_1+\tilde{E}_2, & -2,&0)\cr
\end{array}\right)
\left(\begin{array}{r}
\delta_{\alpha\beta}\cr
\left[U{\cal E}U^{-1}+{\cal A}
\right]_{\alpha\beta}\cr
\left[\left(U{\cal E}U^{-1}+{\cal A}\right)^2\right]_{\alpha\beta}
\end{array}\right)\nonumber\cr
&{\ }&+
\left(\begin{array}{ccc}
\displaystyle
\frac{1}{\Delta \tilde{E}_{21} \Delta \tilde{E}_{31}}
(+\tilde{E}_2\tilde{E}_3, & -(\tilde{E}_2+\tilde{E}_3),&
+1)\cr
\displaystyle
\frac{1}{\Delta \tilde{E}_{21} \Delta \tilde{E}_{32}}
(-\tilde{E}_3\tilde{E}_1, & +(\tilde{E}_3+\tilde{E}_1),&
-1)\cr
\displaystyle
\frac{1}{\Delta \tilde{E}_{31} \Delta \tilde{E}_{32}}
(+\tilde{E}_1\tilde{E}_2, & -(\tilde{E}_1+\tilde{E}_2),&
+1)\cr
\end{array}\right)
\left(\begin{array}{r}
0\cr
\delta_{\alpha\beta}\cr
2\left[U{\cal E}U^{-1}+{\cal A}
\right]_{\alpha\beta}
\end{array}\right)
=0,
\nonumber
\end{eqnarray}
and the terms quadratic in $c$ are given by
\begin{eqnarray}
&{\ }&(V^{-1})^{(2)}\vec{B}^{(0)}+(V^{-1})^{(1)}\vec{B}^{(1)}
+(V^{-1})^{(0)}\vec{B}^{(2)}\nonumber\cr
&=&
\left(\begin{array}{ccc}
\displaystyle
\frac{1}{\Delta \tilde{E}_{21} \Delta \tilde{E}_{31}}
(+1, & 0,&0)\cr
\displaystyle
\frac{1}{\Delta \tilde{E}_{21} \Delta \tilde{E}_{32}}
(-1, & 0,&0)\cr
\displaystyle
\frac{1}{\Delta \tilde{E}_{31} \Delta \tilde{E}_{32}}
(+1, & 0,&0)\cr
\end{array}\right)
\left(\begin{array}{r}
\delta_{\alpha\beta}\cr
\left[U{\cal E}U^{-1}+{\cal A}
\right]_{\alpha\beta}\cr
\left[\left(U{\cal E}U^{-1}+{\cal A}\right)^2\right]_{\alpha\beta}
\end{array}\right)\nonumber\cr
&{\ }&+
\left(\begin{array}{ccc}
\displaystyle
\frac{1}{\Delta \tilde{E}_{21} \Delta \tilde{E}_{31}}
(\tilde{E}_2+\tilde{E}_3, & -2,&0)\cr
\displaystyle
\frac{1}{\Delta \tilde{E}_{21} \Delta \tilde{E}_{32}}
(-(\tilde{E}_3+\tilde{E}_1), & +2,&0)\cr
\displaystyle
\frac{1}{\Delta \tilde{E}_{31} \Delta \tilde{E}_{32}}
(\tilde{E}_1+\tilde{E}_2, & -2,&0)\cr
\end{array}\right)
\left(\begin{array}{r}
0\cr
\delta_{\alpha\beta}\cr
2\left[U{\cal E}U^{-1}+{\cal A}
\right]_{\alpha\beta}
\end{array}\right)\nonumber\cr
&{\ }&+
\left(\begin{array}{ccc}
\displaystyle
\frac{1}{\Delta \tilde{E}_{21} \Delta \tilde{E}_{31}}
(+\tilde{E}_2\tilde{E}_3, & -(\tilde{E}_2+\tilde{E}_3),&
+1)\cr
\displaystyle
\frac{1}{\Delta \tilde{E}_{21} \Delta \tilde{E}_{32}}
(-\tilde{E}_3\tilde{E}_1, & +(\tilde{E}_3+\tilde{E}_1),&
-1)\cr
\displaystyle
\frac{1}{\Delta \tilde{E}_{31} \Delta \tilde{E}_{32}}
(+\tilde{E}_1\tilde{E}_2, & -(\tilde{E}_1+\tilde{E}_2),&
+1)\cr
\end{array}\right)
\left(\begin{array}{r}
0\cr
0\cr
\delta_{\alpha\beta}
\end{array}\right)
=0.
\nonumber
\end{eqnarray}
Thus $\tilde{X}^{\alpha\beta}_j~(j=1,2,3)$ is
independent of $c$, as is claimed.

\section{Derivation of Eq. (\ref{probb})\label{appendix2}}
The matrix (\ref{matrixb}) can be rewritten as
\begin{eqnarray}
{\cal M}
=\frac{1}{2}
\left(\begin{array}{rr}
{\bf 1}&i{\bf 1}\cr
i{\bf 1}&{\bf 1}
\end{array}\right)
\left(\begin{array}{cc}
U{\cal E}U^{-1}+i{\cal B}&0\cr
0&U{\cal E}U^{-1}-i{\cal B}
\end{array}\right)
\left(\begin{array}{rr}
{\bf 1}&-i{\bf 1}\cr
-i{\bf 1}&{\bf 1}
\end{array}\right),
\nonumber
\end{eqnarray}
so the problem of diagonalizing the $6\times6$ matrix (\ref{matrixb})
is reduced to diagonalizing the $3\times3$ matrices
$U{\cal E}U^{-1}\pm i{\cal B}$.
Since we are assuming that $\theta_{13}$ and all the CP phases vanish,
all the matrix elements $U_{\alpha j}$ and
${\cal B}_{\alpha\beta}=-{\cal B}_{\beta\alpha}$
are real, $U{\cal E}U^{-1}\pm i{\cal B}$ can be diagonalized by
a unitary matrix and its complex conjugate:
\begin{eqnarray}
U{\cal E}U^{-1}+ i{\cal B}&=&
\tilde{U}\tilde{{\cal E}}\tilde{U}^{-1}\nonumber\\
U{\cal E}U^{-1}- i{\cal B}&=&
\tilde{U}^\ast\tilde{{\cal E}}(\tilde{U}^\ast)^{-1}.
\nonumber
\end{eqnarray}
Therefore, we can diagonalize ${\cal M}$ by a $6\times6$ unitary
matrix $\tilde{{\cal U}}$ as
\begin{eqnarray}
{\cal M}=\tilde{{\cal U}}\left(\begin{array}{cc}
\tilde{{\cal E}}&0\cr
0&\tilde{{\cal E}}
\end{array}\right)\tilde{{\cal U}}^{-1},
\nonumber
\end{eqnarray}
where
\begin{eqnarray}
\tilde{{\cal U}}=\frac{1}{\sqrt{2}}\left(\begin{array}{rr}
{\bf 1}&-i{\bf 1}\cr
-i{\bf 1}&{\bf 1}
\end{array}\right)\left(\begin{array}{ll}
\tilde{U}&0\cr
0&\tilde{U}^\ast
\end{array}\right)
=
\frac{1}{\sqrt{2}}\left(\begin{array}{rr}
\tilde{U}&{\ }-i\tilde{U}^\ast\cr
-i\tilde{U}&\tilde{U}^\ast
\end{array}\right).
\nonumber
\end{eqnarray}
We note in passing that the reason why diagonalization of
the $6\times6$ matrix is reduced to that of the $3\times3$ matrix is
because the two matrices $U{\cal E}U^{-1}$ and ${\cal B}$ are real.

On the other hand, without a magnetic field
the $6\times6$ unitary
matrix ${\cal U}$ is given by
\begin{eqnarray}
{\cal U}=\left(\begin{array}{ll}
U&0\cr
0&U^\ast
\end{array}\right)=\left(\begin{array}{ll}
U&0\cr
0&U
\end{array}\right),
\nonumber
\end{eqnarray}
where the CP phase $\delta$ has dropped out because $\theta_{13}=0$.
From these we can integrate the equation of motion
and we get the fields at the end point:
\begin{eqnarray}
&{\ }&\left(\begin{array}{l}
\Psi(L)\cr
\Psi^c(L)
\end{array}\right)=\tilde{{\cal U}}(L)
\left(\begin{array}{cc}
e^{-i\Phi}&0\cr
0&e^{-i\Phi}
\end{array}\right)
\tilde{{\cal U}}(0)^{-1}\left(\begin{array}{l}
\Psi(0)\cr
\Psi^c(0)
\end{array}\right)\nonumber\\
&=&\frac{1}{2}
\left(\begin{array}{cc}
Ue^{-i\Phi}\tilde{U}^{-1}+
U^\ast e^{-i\Phi}(\tilde{U}^\ast)^{-1}&
-i(Ue^{-i\Phi}\tilde{U}^{-1}-
U^\ast e^{-i\Phi}(\tilde{U}^\ast)^{-1})\cr
i(Ue^{-i\Phi}\tilde{U}^{-1}-
U^\ast e^{-i\Phi}(\tilde{U}^\ast)^{-1})&
Ue^{-i\Phi}\tilde{U}^{-1}+
U^\ast e^{-i\Phi}(\tilde{U}^\ast)^{-1}
\end{array}\right)
\left(\begin{array}{l}
\Psi(0)\cr
\Psi^c(0)
\end{array}\right)
\nonumber
\end{eqnarray}
where
\begin{eqnarray}
\Phi\equiv \int_0^L\tilde{{\cal E}}(t)\,dt,
\nonumber
\end{eqnarray}
and we have assumed that a large magnetic
field exists at the origin whereas there is no
magnetic field at the end point.
Thus the oscillation probabilities
for the adiabatic transition are give by:
\begin{eqnarray}
P(\nu_\alpha\rightarrow\nu_\beta)=
P(\bar{\nu}_\alpha\rightarrow\bar{\nu}_\beta)&=&
\lim_{L\rightarrow\infty}
\left|\frac{1}{2}\left[
Ue^{-i\Phi}\tilde{U}^{-1}+
U^\ast e^{-i\Phi}(\tilde{U}^\ast)^{-1}
\right]_{\alpha\beta}\right|^2\nonumber\\
&=&\sum_{j=1}^3|U_{\beta j}|^2\left[
\mbox{\rm Re}(\tilde{U}_{\alpha j})\right]^2,
\nonumber\\
P(\bar{\nu}_\alpha\rightarrow\nu_\beta)=
P(\nu_\alpha\rightarrow\bar{\nu}_\beta)&=&
\lim_{L\rightarrow\infty}
\left|\frac{1}{2}\left[
Ue^{-i\Phi}\tilde{U}^{-1}-
U^\ast e^{-i\Phi}(\tilde{U}^\ast)^{-1}
\right]_{\alpha\beta}\right|^2\nonumber\\
&=&\sum_{j=1}^3|U_{\beta j}|^2\left[
\mbox{\rm Im}(\tilde{U}_{\alpha j})\right]^2.
\nonumber
\end{eqnarray}
Hence we obtain the following relation:
\begin{eqnarray}
P(\nu_\alpha\rightarrow\nu_\beta)+
P(\bar{\nu}_\alpha\rightarrow\nu_\beta)=
P(\nu_\alpha\rightarrow\nu_\beta)+
P(\nu_\alpha\rightarrow\bar{\nu}_\beta)=
\sum_{j=1}^3|U_{\beta j}|^2|\tilde{U}_{\alpha j}|^2.
\nonumber
\end{eqnarray}
To get $|\tilde{U}_{\alpha j}|^2$, we need
the explicit expression for the eigenvalues
and the quantity $\tilde{X}^{\alpha\alpha}_j$
in the presence of a magnetic field.
In the following we will subtract $E_1\mbox{\bf 1}$
from the energy matrix ${\cal E}$ because it will
only change the phase of the oscillation amplitude.
For simplicity we will put $\theta_{13}=0$, $\theta_{23}=\pi/4$,
and we will consider the limit $\Delta m^2_{21}\rightarrow0$.
Defining $\Delta E_{jk}\equiv\Delta m^2_{jk}/2E$ and
\begin{eqnarray}
{\cal B}_{\alpha\beta}=B\mu_{\alpha\beta}
\equiv\left(
\begin{array}{ccc}
0&-p&-q\cr
p&0&-r\cr
q&r&0
\end{array}\right),
\nonumber
\end{eqnarray}
we have the eigenvalue equation
\begin{eqnarray}
0&=&|\lambda\mbox{\bf 1}-U({\cal E}-E_1\mbox{\bf 1})U^{-1}-i{\cal B}|\nonumber\\
&=&
\lambda^3-\Delta E_{31}\lambda^2
-(p^2+q^2+r^2)\lambda
+\frac{\Delta E_{31}}{2}(p-q)^2.
\label{cubiceq}
\end{eqnarray}
The three roots of the cubic equation (\ref{cubiceq}) are given by
\begin{eqnarray}
\lambda_1=2R\cos\varphi+\frac{\Delta E_{31}}{3},~
\lambda_2=2R\cos(\varphi+\frac{2}{3}\pi)+\frac{\Delta E_{31}}{3},~
\lambda_3=2R\cos(\varphi-\frac{2}{3}\pi)+\frac{\Delta E_{31}}{3},
\nonumber
\end{eqnarray}
where
\begin{eqnarray}
R&\equiv& [(\Delta E_{31}/3)^2+(p^2+q^2+r^2)/3]^{3/2},\nonumber\\
\varphi&\equiv&(1/3)\cos^{-1}\left
[\{(\Delta E_{31}/3)^3+\Delta E_{31}(p^2+q^2+r^2)/6
-\Delta E_{31}(p-q)^2/4\}/R\right].
\nonumber
\end{eqnarray}
The quantity $\tilde{X}^{\alpha\alpha}_j$ in the presence
of a magnetic field is given by
\begin{eqnarray}
\left(\begin{array}{c}
\tilde{X}^{\alpha\alpha}_1\cr
\tilde{X}^{\alpha\alpha}_2\cr
\tilde{X}^{\alpha\alpha}_3
\end{array}\right)
=\left(\begin{array}{ccc}
\displaystyle
\frac{{\ }1}{\Delta \lambda_{21} \Delta \lambda_{31}}
(\lambda_2\lambda_3, & -(\lambda_2+\lambda_3),&
1)\cr
\displaystyle
\frac{-1}{\Delta \lambda_{21} \Delta \lambda_{32}}
(\lambda_3\lambda_1, & -(\lambda_3+\lambda_1),&
1)\cr
\displaystyle
\frac{{\ }1}{\Delta \lambda_{31} \Delta \lambda_{32}}
(\lambda_1\lambda_2, & -(\lambda_1+\lambda_2),&
1)\cr
\end{array}\right)
\left(\begin{array}{c}
1\cr
Y^{\alpha\alpha}_2\cr
Y^{\alpha\alpha}_3
\end{array}\right),
\label{solxb2}
\end{eqnarray}
where
\begin{eqnarray}
Y^{\alpha\alpha}_2&=&
\left[U({\cal E}-E_1\mbox{\bf 1})U^{-1}+i{\cal B}\right]_{\alpha\alpha}
=\Delta E_{31}X^{\alpha\alpha}_3\nonumber\\
&=&\left\{\begin{array}{l}
0~~~~\,\qquad\qquad(\alpha=e)\cr
\Delta E_{31}/2~~\qquad(\alpha=\mu, \tau)
\end{array}\right.
\label{y2}\\
Y^{\alpha\alpha}_3&=&
\left[\left\{U({\cal E}-E_1\mbox{\bf 1})U^{-1}+i{\cal B}
\right\}^2\right]_{\alpha\alpha}
=(\Delta E_{31})^2X^{\alpha\alpha}_3-({\cal B}^2)_{\alpha\alpha}
\nonumber\\
&=&\left\{\begin{array}{l}
q^2+r^2~~~~~~~\qquad\qquad\qquad(\alpha=e)\cr
r^2+p^2+(\Delta E_{31})^2/2~~\qquad(\alpha=\mu)\cr
p^2+q^2+(\Delta E_{31})^2/2~~\qquad(\alpha=\tau)
\end{array}\right.
.
\label{y3}
\end{eqnarray}
In evaluating $Y^{\alpha\alpha}_j$, we have used the facts
$\theta_{13}=0$, $\theta_{23}=\pi/4$, $\Delta E_{21}=0$,
${\cal B}_{\alpha\beta}=-{\cal B}_{\beta\alpha}$, and that
$U({\cal E}-E_1\mbox{\bf 1})U^{-1}$ is a symmetric matrix.
Using all these results, it is straightforward to obtain the explicit
form for
$P(\nu_\alpha\rightarrow\nu_\beta)+P(\bar{\nu}_\alpha\rightarrow\nu_\beta)$
by plugging the results of Eqs. (\ref{solxb2}), (\ref{y2}), (\ref{y3})
into the following (although calculations are tedious):
\begin{eqnarray}
P(\nu_\alpha\rightarrow\nu_e)+P(\bar{\nu}_\alpha\rightarrow\nu_e)
&=&c_{12}^2\tilde{X}^{\alpha\alpha}_1+s_{12}^2\tilde{X}^{\alpha\alpha}_2
\nonumber\\
P(\nu_\alpha\rightarrow\nu_\beta)+P(\bar{\nu}_\alpha\rightarrow\nu_\beta)
&=&\frac{c_{12}^2}{2}\tilde{X}^{\alpha\alpha}_1
+\frac{s_{12}^2}{2}\tilde{X}^{\alpha\alpha}_2
+\frac{1}{2}\tilde{X}^{\alpha\alpha}_3~~(\beta=\mu, \tau),
\nonumber
\end{eqnarray}
where $s_{12}\equiv\sin\theta_{12}$, $c_{12}\equiv\cos\theta_{12}$.

\section{Derivation of Eq. (\ref{probnp})\label{appendix3}}
The oscillation probability (\ref{probnp}) is obtained 
in two steps.  First we will obtain the eigenvalues
of the matrix (\ref{matrixnp}) with Eq. (\ref{potentialnp}) and
then we will plug the expressions for the eigenvalues
into Eq. (\ref{solx}) with ${\cal A}$ replaced by ${\cal A}_{NP}$
given in Eq. (\ref{potentialnp}).

Let us introduce notations for $3\times3$ hermitian matrices:
\begin{eqnarray}
\lambda_2&\equiv&\left(
\begin{array}{ccc}
0&-i&0\cr
i&0&0\cr
0&0&0
\end{array}\right),\quad
\lambda_5\equiv\left(
\begin{array}{ccc}
0&0&-i\cr
0&0&0\cr
i&0&0
\end{array}\right),\quad
\lambda_7\equiv\left(
\begin{array}{ccc}
0&0&0\cr
0&0&-i\cr
0&i&0
\end{array}\right),\nonumber\\
\lambda_0&\equiv&\left(
\begin{array}{ccc}
1&0&0\cr
0&0&0\cr
0&0&1
\end{array}\right),\quad
\lambda_9\equiv\left(
\begin{array}{ccc}
1&0&0\cr
0&0&0\cr
0&0&-1
\end{array}\right),
\nonumber
\end{eqnarray}
where $\lambda_2$, $\lambda_5$ and $\lambda_7$ are
the standard Gell-Mann matrices whereas
$\lambda_0$ and $\lambda_9$ are the notations which
are defined only in this paper.
Simple calculations show that the matrix ${\cal A}_{NP}$
in Eq. (\ref{potentialnp}) can be rewritten as
\begin{eqnarray}
{\cal A}_{NP}=A\,e^{i\gamma\lambda_9}e^{-i\beta\lambda_5}
\left[\lambda_0\,
\frac{1+\epsilon_{ee}+\epsilon_{\tau\tau}}{2}+
\lambda_9\,\sqrt{\left(\frac{1+\epsilon_{ee}-\epsilon_{\tau\tau}}{2}\right)^2
+|\epsilon_{\mu\tau}|^2}\,\right]
e^{i\beta\lambda_5}e^{-i\gamma\lambda_9},
\label{np1}
\end{eqnarray}
where
\begin{eqnarray}
\beta&\equiv&\frac{1}{2}\tan^{-1}
\frac{2|\epsilon_{e\tau}|^2}{1+\epsilon_{ee}-\epsilon_{\tau\tau}},
\nonumber\\
\gamma&\equiv&\frac{1}{2}\mbox{\rm arg}\,(\epsilon_{e\mu}).
\nonumber
\end{eqnarray}
From Eq. (\ref{np1}) we see that
the two potentially non-zero eigenvalues $\lambda_{e'}$
and $\lambda_{\tau'}$ of the matrix (\ref{potentialnp})
are given by
\begin{eqnarray}
\left(\begin{array}{c}
\lambda_{e'}\cr
\lambda_{\tau'}
\end{array}\right)=
A\left[
\frac{1+\epsilon_{ee}+\epsilon_{\tau\tau}}{2}\pm
\sqrt{\left(\frac{1+\epsilon_{ee}-\epsilon_{\tau\tau}}{2}\right)^2
+|\epsilon_{\mu\tau}|^2}\,
\right].
\nonumber
\end{eqnarray}
In order for this scheme to be consistent with
the atmospheric neutrino data particularly at high energy,
which are perfectly described by vacuum oscillations,
$\lambda_{\tau'}$ has to vanish~\cite{Friedland:2005vy}.
In this case, we have
\begin{eqnarray}
\tan\beta&=&\frac{|\epsilon_{e\tau}|}
{1+\epsilon_{ee}},\nonumber\\
\epsilon_{\tau\tau}&=&\frac{|\epsilon_{e\tau}|^2}
{1+\epsilon_{ee}},\nonumber\\
\lambda_{e'}&=&A(1+\epsilon_{ee})
\left[1+\frac{|\epsilon_{e\tau}|^2}
{(1+\epsilon_{ee})^2}\right]=
\frac{A(1+\epsilon_{ee})}{\cos^2\beta}.
\nonumber
\end{eqnarray}
Thus we have
\begin{eqnarray}
{\cal A}_{NP}=A\,e^{i\gamma\lambda_9}e^{-i\beta\lambda_5}
\mbox{\rm diag}\left(\lambda_{e'},0,0\right)
e^{i\beta\lambda_5}e^{-i\gamma\lambda_9}.
\label{np2}
\end{eqnarray}
If we did not have $\beta$ and $\gamma$, Eq. (\ref{np2}) would be the same
as the standard three flavor scheme in matter, which was analytically
worked out in Ref.~\cite{Yasuda:1998sf} in the limit of $\Delta
m^2_{21}\rightarrow0$.
It turns out that,
by redefining the parametrization of the MNS matrix
Eq. (\ref{np2}) can be also treated analytically in the limit of $\Delta
m^2_{21}\rightarrow0$ as was done in Ref.~\cite{Yasuda:1998sf}.
The mass matrix can be written as
\begin{eqnarray}
U{\cal E}U^{-1}+{\cal A}_{NP}
=e^{i\gamma\lambda_9}e^{-i\beta\lambda_5}
\left[
e^{i\beta\lambda_5}e^{-i\gamma\lambda_9}U{\cal E}U^{-1}
e^{i\gamma\lambda_9}e^{-i\beta\lambda_5}
+\mbox{\rm diag}\left(\lambda_{e'},0,0\right)
\right]
e^{i\beta\lambda_5}e^{-i\gamma\lambda_9}.
\nonumber
\end{eqnarray}
Here we introduce the following two unitary matrices:
\begin{eqnarray}
U'&\equiv& e^{i\beta\lambda_5}e^{-i\gamma\lambda_9}\,U\nonumber\\
&\equiv&\mbox{\rm diag}(1,1,e^{i\text{arg}\,U'_{\tau3}})\,
U''\,\mbox{\rm diag}
(e^{i\text{arg}\,U'_{e1}},e^{i\text{arg}\,U'_{e2}},1),
\nonumber
\end{eqnarray}
where $U$ is the $3\times3$ MNS matrix in the standard
parametrization~\cite{Eidelman:2004wy} and
$U''$ was defined in the second line in such a way
that the elements $U''_{e1}$, $U''_{e2}$, $U''_{\tau3}$
be real to be consistent with the standard parametrization
in Ref.~\cite{Eidelman:2004wy}~\footnote{
The element $U''_{\tau2}$ has to be also real, but it is
already satisfied because $U''_{\tau2}=U_{\tau2}$.}.
Then we have
\begin{eqnarray}
U{\cal E}U^{-1}+{\cal A}_{NP}
&=&e^{i\gamma\lambda_9}e^{-i\beta\lambda_5}
\mbox{\rm diag}(1,1,e^{i\text{arg}\,U'_{\tau3}})\,
\left[U''{\cal E}U''^{-1}
+\mbox{\rm diag}\left(\lambda_{e'},0,0\right)
\right]\nonumber\\
&{\ }&\times\mbox{\rm diag}(1,1,e^{-i\text{arg}\,U'_{\tau3}})\,
e^{i\beta\lambda_5}e^{-i\gamma\lambda_9}.
\label{np3}
\end{eqnarray}
Before proceeding further, let us obtain the
expression for the three mixing angles $\theta''_{jk}$
and the Dirac phase $\delta''$ in $U''$.
Since
\begin{eqnarray}
U'=\left(\begin{array}{ccc}
c_\beta e^{-i\gamma}U_{e1}+s_\beta e^{i\gamma}U_{\tau1}&
c_\beta e^{-i\gamma}U_{e2}+s_\beta e^{i\gamma}U_{\tau2}&
c_\beta e^{-i\gamma}U_{e3}+s_\beta e^{i\gamma}U_{\tau3}\cr
U_{\mu1}&U_{\mu2}&U_{\mu3}\cr
c_\beta e^{-i\gamma}U_{\tau1}-s_\beta e^{i\gamma}U_{e1}&
c_\beta e^{-i\gamma}U_{\tau2}-s_\beta e^{i\gamma}U_{e2}&
c_\beta e^{-i\gamma}U_{\tau3}-s_\beta e^{i\gamma}U_{e3}
\end{array}
\right),
\nonumber
\end{eqnarray}
where $c_\beta\equiv\cos\beta$, $s_\beta\equiv\sin\beta$,
we get
\begin{eqnarray}
\theta''_{13}&=&\sin^{-1}|U''_{e3}|=
\sin^{-1}|c_\beta e^{-i\gamma}U_{e3}+s_\beta e^{i\gamma}U_{\tau3}|
\nonumber\\
\theta''_{12}&=&\tan^{-1}(U''_{e2}/U''_{e1})=
\tan^{-1}\left(|c_\beta e^{-i\gamma}U_{e2}+s_\beta e^{i\gamma}U_{\tau2}|
/|c_\beta e^{-i\gamma}U_{e1}+s_\beta e^{i\gamma}U_{\tau1}|\right)
\nonumber\\
\theta''_{23}&=&\tan^{-1}(U''_{\mu3}/U''_{\tau3})=
\tan^{-1}\left(U_{\mu3}/
|c_\beta e^{-i\gamma}U_{\tau3}-s_\beta e^{i\gamma}U_{e3}|\right)
\nonumber\\
\delta''&=&-\mbox{\rm arg}\,U''_{e3}=
-\mbox{\rm arg}\,(c_\beta e^{-i\gamma}U_{e3}+s_\beta e^{i\gamma}U_{\tau3}).
\nonumber
\end{eqnarray}
As was shown in Ref.~\cite{Yasuda:1998sf}, in the limit
$\Delta m^2_{21}\rightarrow0$, the matrix on the right hand side
of Eq. (\ref{np3}) can be diagonalized as follows:
\begin{eqnarray}
&{\ }&U''{\cal E}U''^{-1}
+\mbox{\rm diag}\left(\lambda_{e'},0,0\right)-E_1{\bf 1}\nonumber\\
&=&e^{i\theta''_{23}\lambda_7}\Gamma_{\delta''}
e^{i\theta''_{13}\lambda_5}\Gamma_{\delta''}^{-1}
e^{i\theta''_{12}\lambda_2}
\mbox{\rm diag}\left(0,0,\Delta E_{31}\right)
e^{-i\theta''_{12}\lambda_2}\Gamma_{\delta''}
e^{-i\theta''_{13}\lambda_5}\Gamma_{\delta''}^{-1}
e^{-i\theta''_{23}\lambda_7}
+\mbox{\rm diag}\left(\lambda_{e'},0,0\right)\nonumber\\
&=&e^{i\theta''_{23}\lambda_7}\Gamma_{\delta''}
\left[e^{i\theta''_{13}\lambda_5}\mbox{\rm diag}\left(0,0,\Delta E_{31}\right)
+\mbox{\rm diag}\left(\lambda_{e'},0,0\right)\right]
\Gamma_{\delta''}^{-1}e^{-i\theta''_{23}\lambda_7}\nonumber\\
&=&e^{i\theta''_{23}\lambda_7}\Gamma_{\delta''}
e^{i\tilde{\theta}''_{13}\lambda_5}
\mbox{\rm diag}\left(\Lambda_-,0,\Lambda_+\right)
e^{-i\tilde{\theta}''_{13}\lambda_5}
\Gamma_{\delta''}^{-1}e^{i\theta''_{12}\lambda_2},
\nonumber
\end{eqnarray}
where $\Gamma_{\delta''}\equiv\mbox{\rm diag}(1,1,e^{-i\delta''})$,
$\Delta E_{31}\equiv\Delta m_{31}^2/2E$,
we have used the standard parametrization~\cite{Eidelman:2004wy}
$U''\equiv e^{i\theta''_{23}\lambda_7}\Gamma_{\delta''}
e^{i\theta''_{13}\lambda_5}\Gamma_{\delta''}^{-1}
e^{i\theta''_{12}\lambda_2}$, and the eigenvalues
$\Lambda_\pm$ are defined by
\begin{eqnarray}
\Lambda_\pm&=&\frac{1}{2}\left(\Delta E_{31}+\lambda_{e'}
\right)
\pm\frac{1}{2}
\sqrt{\left(\Delta E_{31}\cos2\theta_{13}''
-\lambda_{e'}\right)^2
+(\Delta E_{31}\sin2\theta_{13}'')^2}.
\nonumber
\end{eqnarray}

Having obtained the eigenvalues, by plugging these into Eq. (\ref{solx})
with ${\cal A}\rightarrow{\cal A}_{NP}$,
$\tilde{E}_1\rightarrow\Lambda_-$,
$\tilde{E}_2\rightarrow0$,
$\tilde{E}_3\rightarrow\Lambda_+$, we obtain
$\tilde{X}^{\mu e}$:
\begin{eqnarray}
\left(\begin{array}{c}
\tilde{X}^{\mu e}_1\cr
\tilde{X}^{\mu e}_2\cr
\tilde{X}^{\mu e}_3
\end{array}\right)
=\left(\begin{array}{ccc}
\displaystyle
\frac{-1}{\Lambda_-(\Lambda_+-\Lambda_-)}
(0, & -\Lambda_+,&1)\cr
\displaystyle
\frac{{\ }1}{\Lambda_+\Lambda_-}
(-\Lambda_+\Lambda_-, & -(\Lambda_++\Lambda_-),&
1)\cr
\displaystyle
\frac{{\ }1}{\Lambda_+(\Lambda_+-\Lambda_-)}
(0, & -\Lambda_-,&1)\cr
\end{array}\right)
\left(\begin{array}{r}
0\cr
Y^{\mu e}_2\cr
Y^{\mu e}_3
\end{array}\right)=
\left(\begin{array}{c}
\displaystyle
\frac{-Y^{\mu e}_3+\Lambda_+Y^{\mu e}_2}{\Lambda_-(\Lambda_+-\Lambda_-)}\cr
\displaystyle
\frac{Y^{\mu e}_3-(\Lambda_++\Lambda_-)Y^{\mu e}_2}{\Lambda_+\Lambda_-}\cr
\displaystyle
\frac{Y^{\mu e}_3-\Lambda_-Y^{\mu e}_2}{\Lambda_+(\Lambda_+-\Lambda_-)}
\end{array}\right),
\nonumber
\end{eqnarray}
where $Y^{\mu e}_j$ are defined by
\begin{eqnarray}
Y^{\mu e}_j\equiv\left[
\left(U{\cal E}U^{-1}+{\cal A}_{NP}\right)^{j-1}\right]_{\mu e},
\nonumber
\end{eqnarray}
and are given by
\begin{eqnarray}
Y^{\mu e}_2&=&\Delta E_{31}\,X^{\mu e}_3\nonumber\\
Y^{\mu e}_3&=&[(\Delta E_{31})^2+A(1+\epsilon_{ee})\Delta E_{31}]
X^{\mu e}_3+A\Delta E_{31}\epsilon_{e\tau}^\ast\,X^{\mu \tau}_3.
\nonumber
\end{eqnarray}
Furthermore, by introducing the notations
\begin{eqnarray}
\xi&\equiv&[(\Delta E_{31})^2+A(1+\epsilon_{ee})
\Delta E_{31}]U_{\mu3}|U_{e3}|\nonumber\\
\eta&\equiv&A\Delta E_{31}|\epsilon_{e\tau}|U_{\mu3}U_{\tau3}
\nonumber\\
\zeta&\equiv&\Delta E_{31}U_{\mu3}|U_{e3}|,
\nonumber
\end{eqnarray}
we can rewrite $Y^{\mu e}_2=\zeta e^{i\delta}$ and
$Y^{\mu e}_3=\xi e^{i\delta}+\eta e^{-2i\gamma}$,
where $\delta$ is the Dirac CP phase of the MNS matrix $U$,
so we have
\begin{eqnarray}
\tilde{X}^{\mu e}_1&=&
\frac{-e^{i\delta}}{\Lambda_-(\Lambda_+-\Lambda_-)}
[\xi+\eta e^{-i(2\gamma+\delta)}-\Lambda_+\zeta]\nonumber\\
\tilde{X}^{\mu e}_2&=&
\frac{e^{i\delta}}{\Lambda_+\Lambda_-}
[\xi+\eta e^{-i(2\gamma+\delta)}-(\Lambda_++\Lambda_-)\zeta]\nonumber\\
\tilde{X}^{\mu e}_3&=&
\frac{e^{i\delta}}{\Lambda_+(\Lambda_+-\Lambda_-)}
[\xi+\eta e^{-i(2\gamma+\delta)}-\Lambda_-\zeta].
\nonumber
\end{eqnarray}
Notice that the phase factor $e^{i\delta}$ in front of each $\tilde{X}^{\mu e}_j$
drops out in the oscillation probability $P(\nu_\mu\rightarrow\nu_e)$
because $P(\nu_\mu\rightarrow\nu_e)$ is expressed
in terms of $\tilde{X}^{\mu e}_j\tilde{X}^{\mu e\ast}_k$, and
the oscillation probability (\ref{probnp})
depends only on the combination
$2\gamma+\delta=\mbox{\rm arg}\,(\epsilon_{e\mu})+\delta$.

In the present case, the matrix $\tilde{U}$ is unitary
and because of this three flavor unitarity all the T violating terms
are proportional to one factor:
\begin{eqnarray}
&{\ }&2\sum_{j<k}\mbox{\rm Im}\left(\tilde{X}^{\mu e}_j
\tilde{X}^{\mu e\ast}_k\right)
\sin\left(\Delta \tilde{E}_{jk}L\right)\nonumber\\
&=&
2\,\mbox{\rm Im}\left(\tilde{X}^{\mu e}_1\tilde{X}^{\mu e\ast}_2\right)
[\sin\left(\Delta \tilde{E}_{12}L\right)
-\sin\left(\Delta \tilde{E}_{13}L\right)
+\sin\left(\Delta \tilde{E}_{23}L\right)]\nonumber\\
&=&-8\,\mbox{\rm Im}\left(\tilde{X}^{\mu e}_1\tilde{X}^{\mu e\ast}_2\right)
\sin\left(\frac{\Delta \tilde{E}_{21}L}{2}\right)
\sin\left(\frac{\Delta \tilde{E}_{31}L}{2}\right)
\sin\left(\frac{\Delta \tilde{E}_{32}L}{2}\right).
\nonumber
\end{eqnarray}
This modified Jarlskog factor
$\mbox{\rm Im}(\tilde{X}^{\mu e}_1\tilde{X}^{\mu e\ast}_2)$
in matter can be rewritten as
\begin{eqnarray}
\mbox{\rm Im}(\tilde{X}^{\mu e}_1\tilde{X}^{\mu e\ast}_2)
&=&
\frac{1}{\Lambda_+\Lambda_-(\Lambda_+-\Lambda_-)}
\mbox{\rm Im}(Y^{\mu e}_3Y^{\mu e\ast}_2)
=
-\frac{\eta\zeta\sin(2\gamma+\delta)}
{\Lambda_+\Lambda_-(\Lambda_+-\Lambda_-)}
\nonumber\\
&=&
-\frac{A(\Delta E_{31})^2}{\Lambda_+\Lambda_-(\Lambda_+-\Lambda_-)}
\,\left|\epsilon_{e\tau}X^{e\mu}_3X^{\mu \tau}_3\right|\,
\sin(\mbox{\rm arg}(\epsilon_{e\mu})+\delta).
\nonumber
\end{eqnarray}
This completes derivation of Eq. (\ref{probnp}).

\section*{Acknowledgments}
The author would like to thank Alexei Smirnov for bringing my attention
to Refs.~\cite{Grimus:1993fz,Halprin:1986pn,Mannheim:1987ef,Sawyer:1990tw}.
He would also like to thank He Zhang for calling my attention to
Refs.~\cite{Xing:2005gk,Zhang:2006yq} which were missed in the
first version of this paper.
This research was supported in part by a Grant-in-Aid for Scientific
Research of the Ministry of Education, Science and Culture,
\#19340062.

\end{document}